\documentclass[twocolumn,showpacs,preprintnumbers,amsmath,amssymb]{revtex4}

\usepackage{graphicx}
\usepackage{dcolumn}
\usepackage{bm}

\begin{document}

\title{\textbf{Velocity-locked solitary waves in quadratic media} }

\author{Fabio Baronio, Matteo Conforti, Costantino De Angelis}
\affiliation{ CNISM, Universit\`a di Brescia, Via Branze 38, 25123
Brescia, Italy
}%

\author{Antonio Degasperis}
\affiliation{Dipartimento di Fisica and INFN, Universit\`a
``La Sapienza'', P.le A. Moro 2, 00185 Roma, Italy}%

\author{Marco Andreana, Vincent Couderc, Alain Barth\'el\'emy}
\affiliation{XLIM, CNRS and Universit\'e de
Limoges, Av. Albert Thomas 123, 87060, Limoges, France}%

\date{\today}

\begin{abstract}
\textbf{We demonstrate experimentally the existence of three-wave
resonant interaction solitary triplets in quadratic media.} Stable
velocity-locked bright-dark-bright spatial solitary triplets,
determined by the balance between the energy exchange rates and
the velocity mismatch between the interacting waves, are excited
in a KTP crystal.
%
\end{abstract}

\maketitle

In recent years \textbf{solitary waves} in quadratic materials
have been the subject of an intense renewal of interest from both
theoretical and experimental viewpoints. Two types of
\textbf{solitary waves} that were both predicted in the early
1970's are being studied. On one hand, one finds solitary waves
that result from a balance between nonlinearity and diffraction
(or dispersion) \cite{kara74}. This type of \textbf{solitary wave}
has been intensively investigated experimentally over the past few
years \cite{tril02}. On the other hand, quadratic media were shown
to support solitary waves that result from energy exchanges
between diffractionless (or dispersionless) waves of different
velocities \cite{armstrong70,nozaki73,kaup79}. The structure of
\textbf{these solitary} waves is determined by the balance between
the energy exchange rates and the velocity mismatch between the
interacting waves \cite{tril96,dega06}. This type of solitary wave
is ubiquitous in nonlinear wave systems \cite{kaup79} and has been
reported in such diverse fields as plasma physics, hydrodynamics,
acoustics, and nonlinear optics, in particular in the context of
self-induced transparency \cite{nozaki73,mcca67}. This type of
wave has also been investigated experimentally in stimulated Raman
scattering in gases \cite{druh83} and recently in $H_2$-filled
photonic crystal fibers \cite{russ09}, in stimulated Brillouin
fiber-ring lasers \cite{pich91}, but no experiments have been
reported to date on \textbf{solitary waves} of quadratic optical
materials.

In this Letter we report the experimental observation of
diffractionless \textbf{velocity-locked solitary triplets} in a
quadratic crystal. We consider the optical spatial non-collinear
scheme with type II second harmonic generation (SHG) in a KTP
crystal. A spatial narrow diffractionless extraordinary beam (the
signal) and an ordinary quasi-plane wave (the pump), both at the
fundamental frequency (FF) $\omega$, mix via $\chi^{(2)}$ to
generate a second harmonic (SH) beam at frequency $2\omega$ (the
idler). Depending on the input intensities, three different
regimes exist. Linear regime: the FF signal and pump beams do not
interact. Frequency conversion: the FF signal and pump beams
interact and generate a SH idler whose spatial characteristics are
associated with the interaction distance in the crystal; signal
and pump are depleted. Solitary regime: the FF signal and pump
beams interact, generate a spatial narrow SH idler, a spatial dip
appears in the pump, whereas the intensity and propagation
direction of the signal beam are modified. Indeed, the interaction
generates a \textbf{stable bright-dark-bright triplet} moving with
a locked spatial nonlinear velocity \cite{dega06}.

\begin{figure}[h]
\begin{center}
\includegraphics[width=7.8cm]{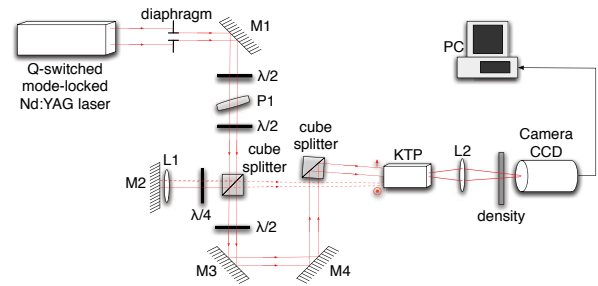}
    \end{center}
     \caption{Experimental set-up.
     M1, M2, M3, M4: mirrors. P1: polarizer. L1, L2: lenses.
    } \label{esperimento}
\end{figure}
\begin{figure}[h]
\begin{center}
\includegraphics[width=6cm]{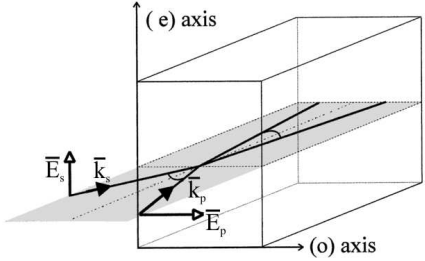}
    \end{center}
     \caption{Schematic representation the optical non collinear configuration in the KTP crystal.}
     \label{schema}
\end{figure}

In the experiments (see Fig. \ref{esperimento}), a Q–-switched,
mode–-locked Nd:YAG laser delivers $40 ps$ pulses at $\lambda=1064
nm$. We introduce a Glan polarizer to obtain, after passage of the
light through $P_1$, two independent beams with perpendicular
linear polarization states. A half–-wave plate placed before the
prism serves to adjust the intensity of the two beams. By means of
highly reflecting mirrors, beam splitters and lenses the beams are
focused and spatially superimposed in the plane of their beam
waist with a circular shape of $200 \mu m$ and $2.2 mm$, full
width at half maximum in intensity, for the signal and pump waves
respectively. A $L=3 cm$ long type II KTP crystal cut for second
harmonic generation is positioned such that its input face
corresponds to the plane of superposition of the two input beams.
The crystal is oriented for perfect phase matching. The directions
of the linear polarization state of the two beams are adjusted to
coincide with the extraordinary and the ordinary axes,
respectively, of the KTP crystal. The wave vectors of the input
beams are tilted at angles of $\theta_s=2.1^o$ and
$\theta_p=-2.1^o$ (in the crystal) with respect to the direction
of perfect collinear phase matching for the extraordinary and the
ordinary components, respectively (see Fig. \ref{schema}). These
parameters correspond inside the crystal to a tilt between the
input beams greater than the natural walk-off angle but introduced
along the ordinary noncritical plane. The idler second harmonic
direction lies in between the input beams directions ($\theta_i
\cong 0.4^o$). With these values of parameters, spatial
diffraction and temporal dispersion were negligible. The spatial
waves' patterns at the output of the crystal are imaged with
magnification onto a CCD camera and analyzed. We use alternately
different filters and polarizers to select either the IR or the
green output.

\begin{figure}[h]
\begin{center}
     \includegraphics[width=6cm]{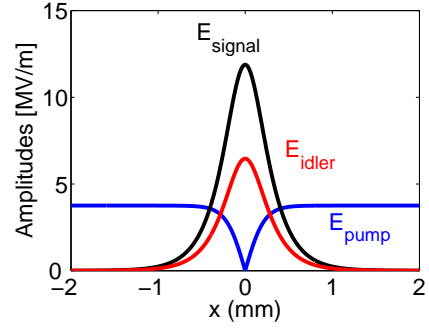}
\end{center}
     \caption{Envelopes $E_s$, $E_p$ and $E_i$ of the \textbf{solitary
     triplet.}
    } \label{inviluppi}
\end{figure}

Theoretically, the equations describing the spatial quadratic
resonant interaction of three waves in the quadratic nonlinear
medium read as:

\begin{eqnarray}\label{3wri}
\nonumber \Big ( \frac{\partial}{\partial z}-\rho_s \frac{\partial
}{\partial x} \Big ) E_s + \frac{1}{2 i k_s} \Big (
\frac{\partial^2}{\partial x^2} + \frac{\partial^2}{\partial y^2}
\Big ) E_s &=& i \chi_s E_p^*E_i,\\ \nonumber \Big (
\frac{\partial}{\partial z}-\rho_p \frac{\partial }{\partial x}
\Big ) E_p + \frac{1}{2 i k_p} \Big ( \frac{\partial^2}{\partial
x^2} + \frac{\partial^2}{\partial y^2} \Big ) E_p &=& i \chi_p
E_s^*E_i,\\ \nonumber \Big ( \frac{\partial}{\partial z}-\rho_i
\frac{\partial }{\partial x} \Big ) E_i + \frac{1}{2 i k_i} \Big (
\frac{\partial^2}{\partial x^2} + \frac{\partial^2}{\partial y^2}
\Big ) E_i &=& i \chi_i E_s E_p.\\
\end{eqnarray}
$E_j(x,y,z)$ are the slowly varying electric field envelopes of
the waves at frequencies $\omega_j$ (wavelength $\lambda_j$),
$k_j=\omega_j n_j / c$ are the wavenumbers, $n_j$ the refractive
indexes, $\chi_j=2 d \omega_j / c n_j$ the nonlinear coupling
coefficients ($d$ is the quadratic nonlinear susceptibility and
$c$ is the speed of light), $\rho_j$ the walk off angles and
$j=s,p,i$ (s:signal, p:pump, i:idler). $z$ is the spatial
longitudinal propagation coordinate, $x$ and $y$ are the spatial
transverse coordinates. In the configuration we considered
$\rho_p<\rho_i<\rho_s$, spatial diffraction is negligible;
therefore, Eqs. (\ref{3wri}) reduce to the integrable three-wave
model reported in Ref. \cite{dega06}, and in the ordinary $x-z$
plane ($y=0$ plane) \textbf{ bright-dark-bright triplets }that
travel with a common nonlinear locked velocity can be excited.
Typical envelope profiles of the \textbf{bright-dark-bright}
solitary waves of Eqs. (\ref{3wri}) versus $x$ for fixed $z$ and
$y$ are shown in Fig. \ref{inviluppi}.

As the intensities of the input signal and pump are varied in a
suitable range, we observed three different regimes in the
ordinary KTP  $x-z$ ($y=0$) plane: linear, frequency conversion,
and solitary regimes.

\begin{figure}[h]
\begin{center}
     \includegraphics[width=2.7cm]{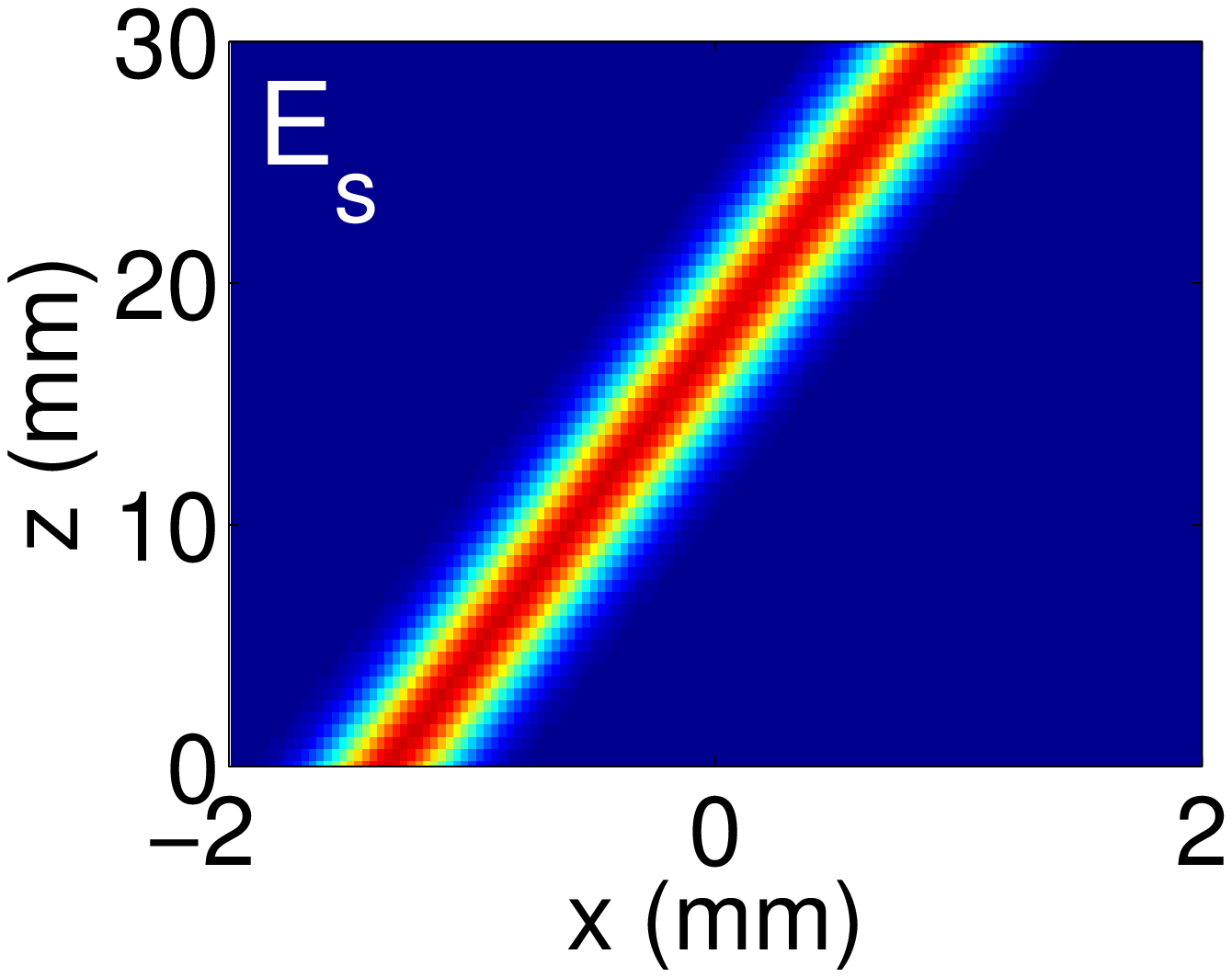}, \includegraphics[width=2.7cm]{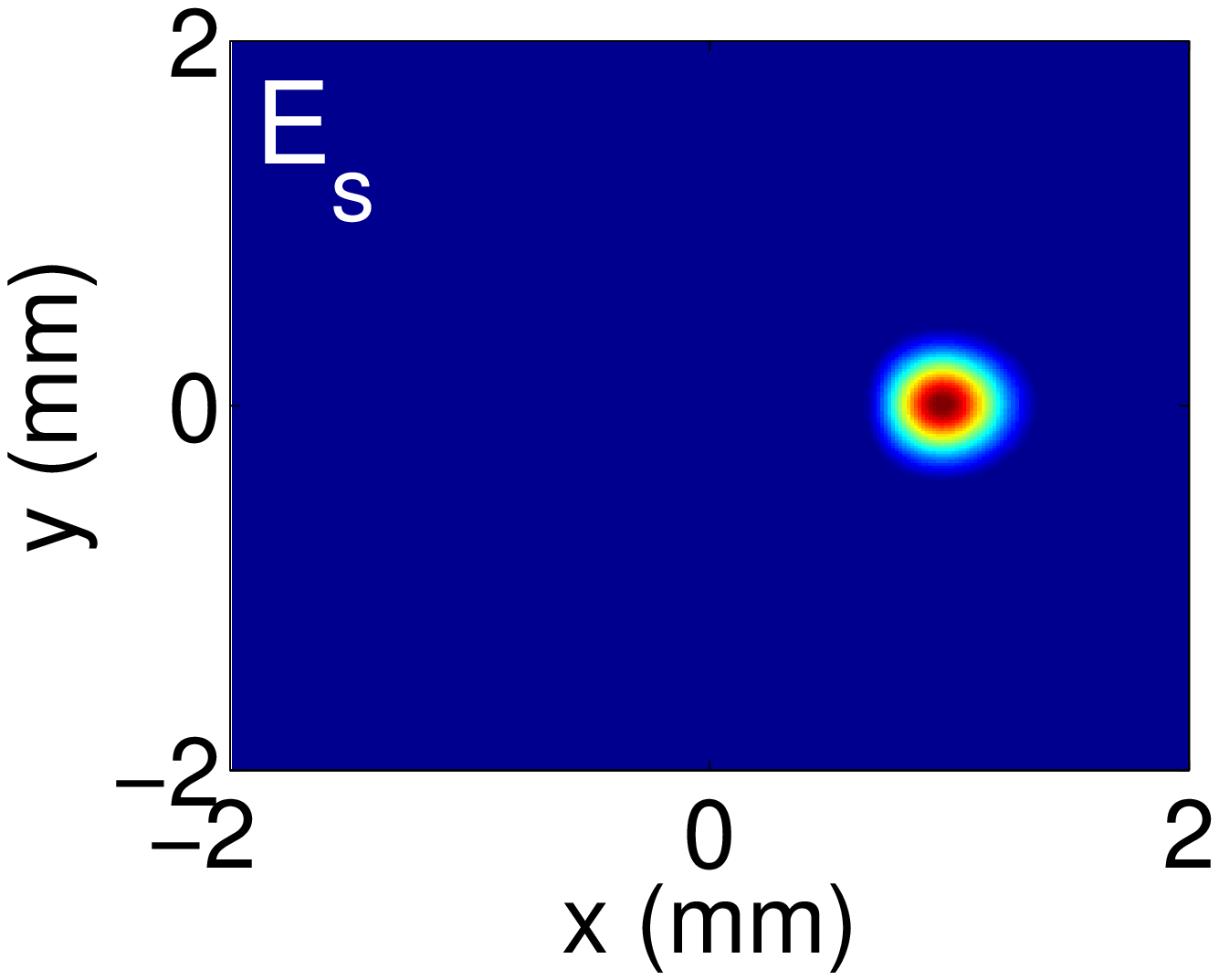}, \includegraphics[width=2.7cm]{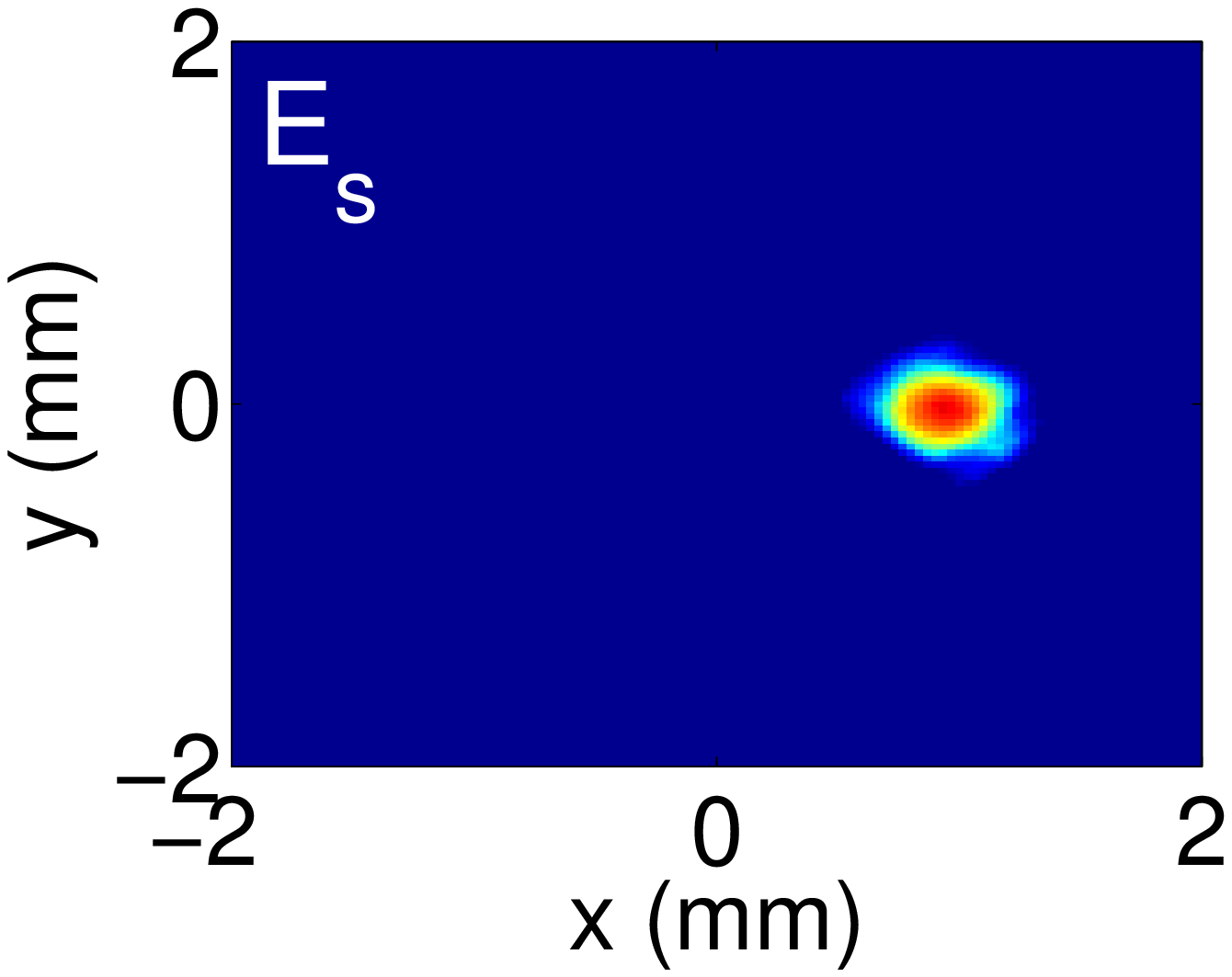} \\
     \includegraphics[width=2.7cm]{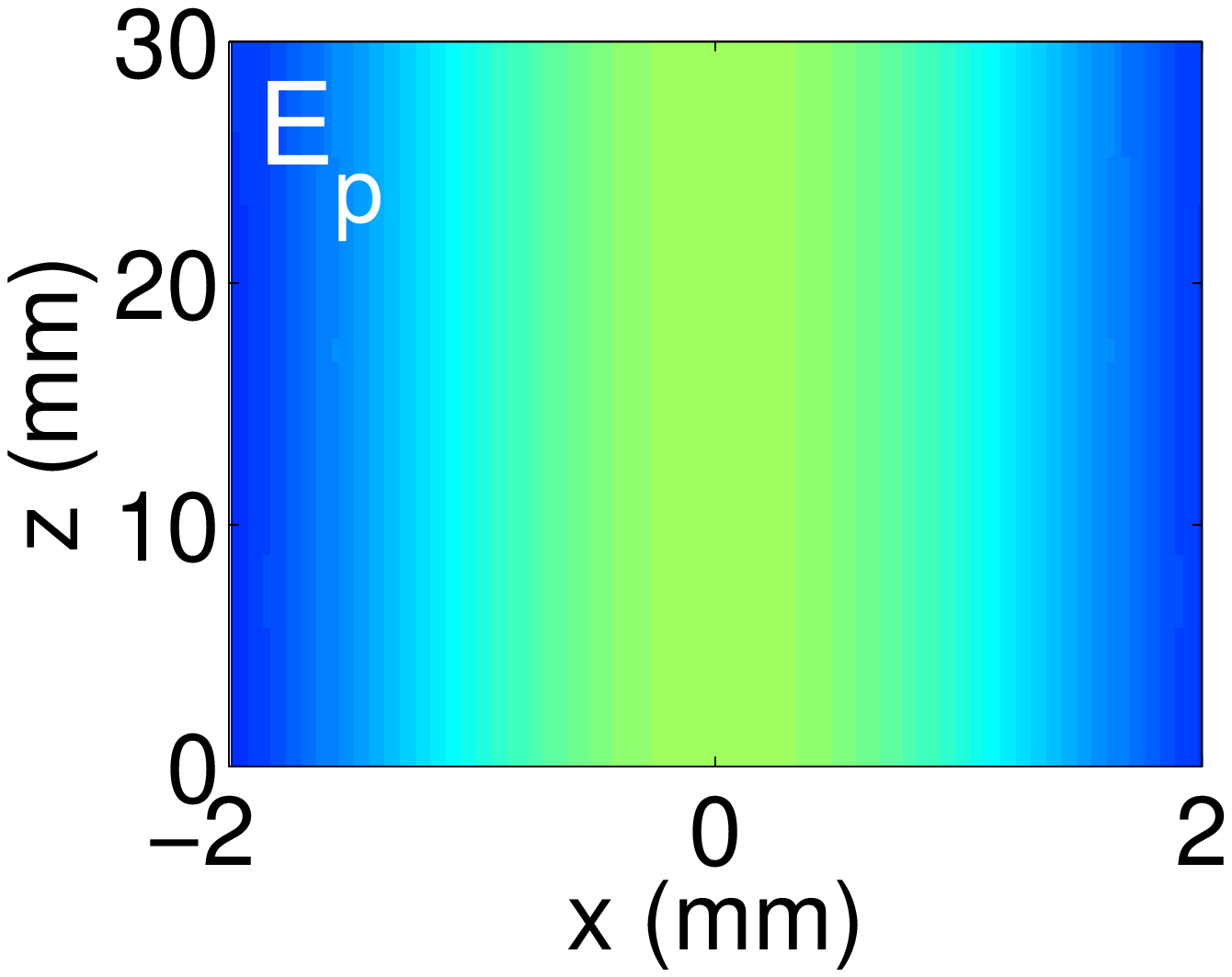}, \includegraphics[width=2.7cm]{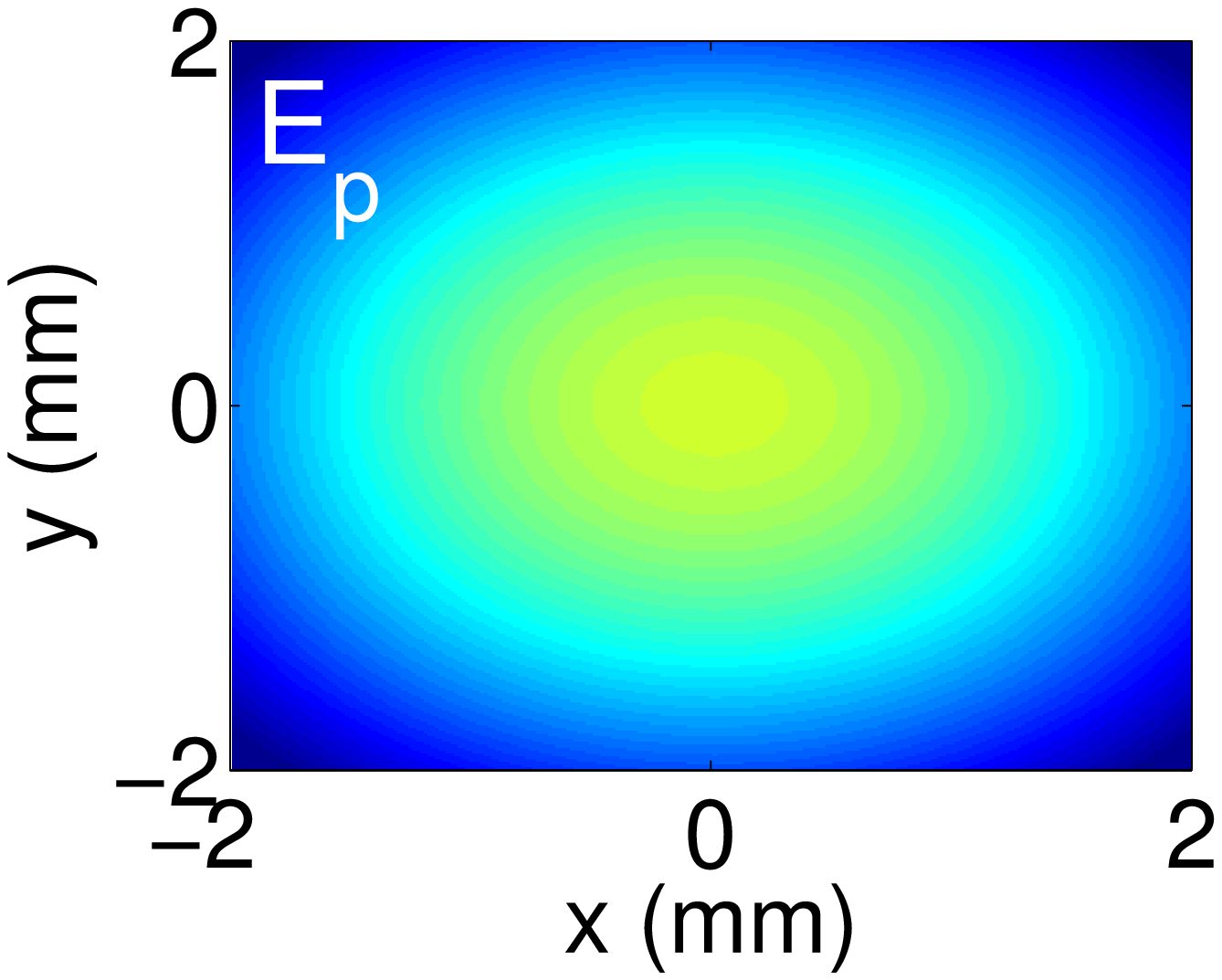}, \includegraphics[width=2.7cm]{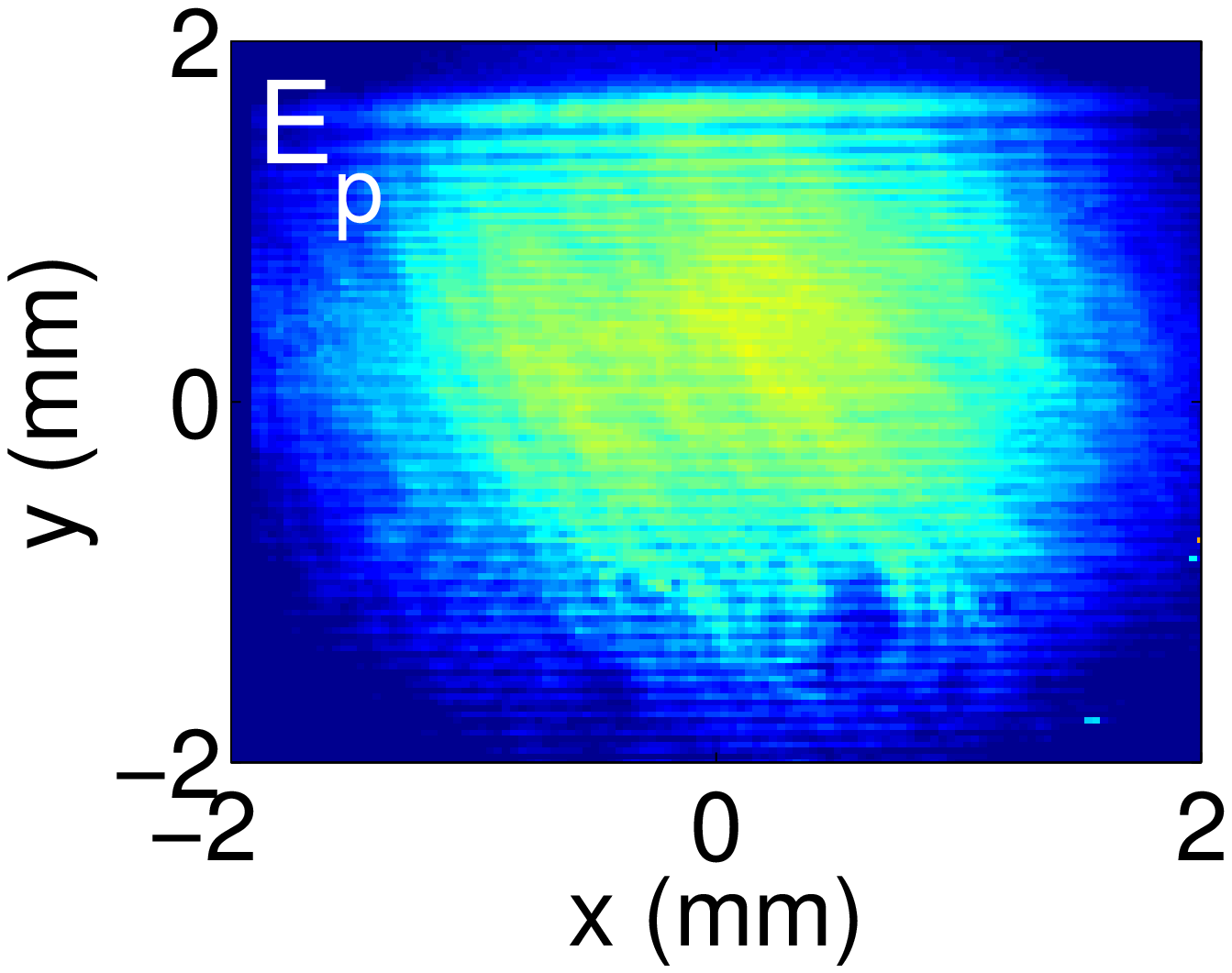} \\
\end{center}
     \caption{Linear regime. Left column, numerical dynamics of the beams $E_s$, $E_p$ in the
     $x-z$ ($y=0$) plane. Central column, numerical, and right column, experimental
     results at the exit face of the KTP crystal presenting
     the spatial $x-y$ output profiles of the beams.
     At the input $I_s=0.1 MW/cm^2$, $I_p=0.01 MW/cm^2$.
    } \label{lineare}
\end{figure}

In the low-intensity linear regime ($I_s=0.1 MW/cm^2$, $I_p=0.01
MW/cm^2$), the signal and the pump do not interact and propagate
without diffraction in the KTP crystal following their own
characteristic spatial directions (Fig. \ref{lineare}). Left
column of Fig. \ref{lineare} shows the numerical spatial evolution
of the extraordinary polarized signal and the ordinary polarized
pump in the ordinary $x-z$ ($y=0$) plane; central and right
columns report respectively the numerical and experimental spatial
output profiles of the beams in the $x-y$ ($z=L$) plane. The
numerical and experimental results are reported considering a
spatial frame moving with the pump walk off angle. No SH
extraordinary polarized idler is generated.

\begin{figure}[h]
\begin{center}
         \includegraphics[width=2.7cm]{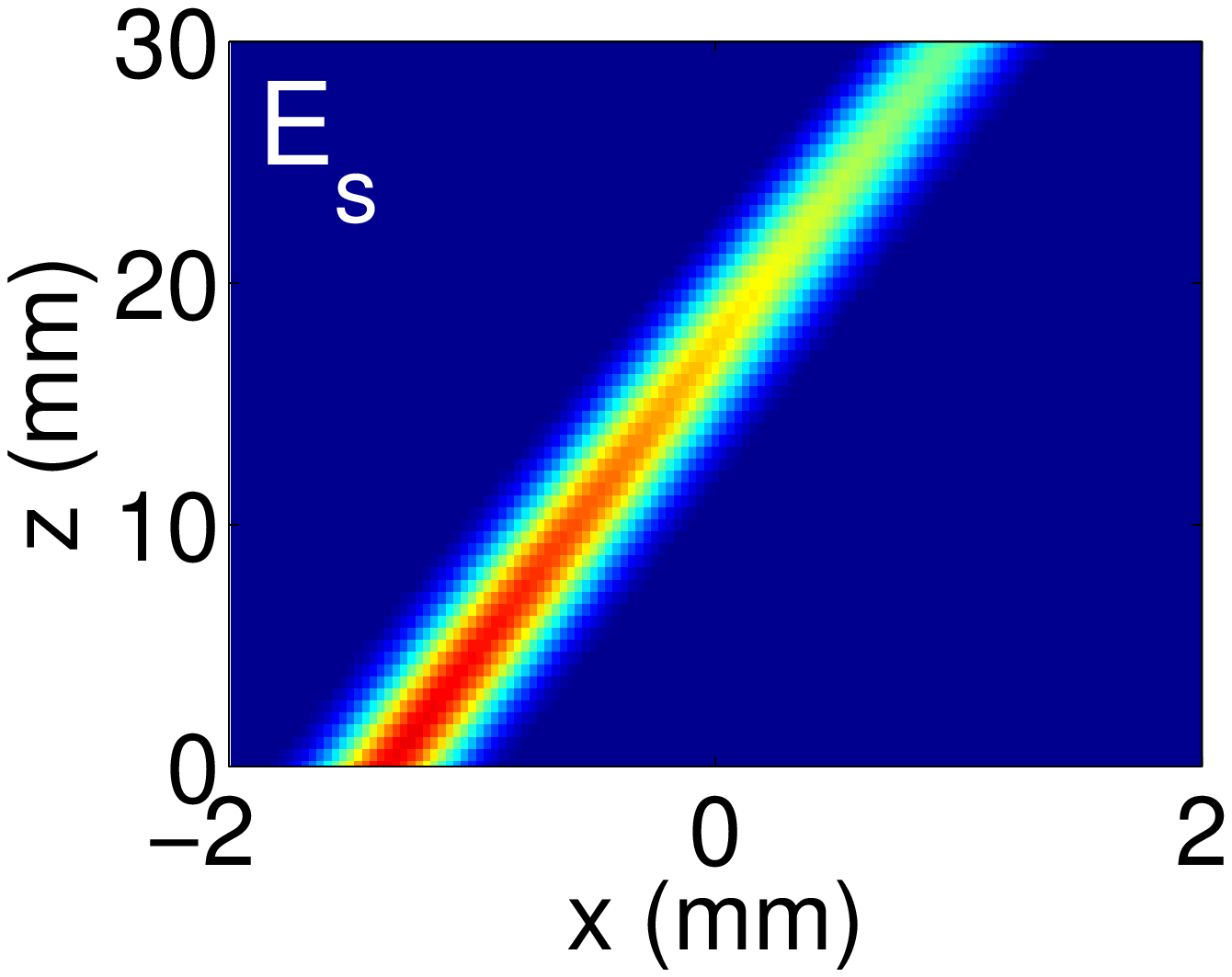}, \includegraphics[width=2.7cm]{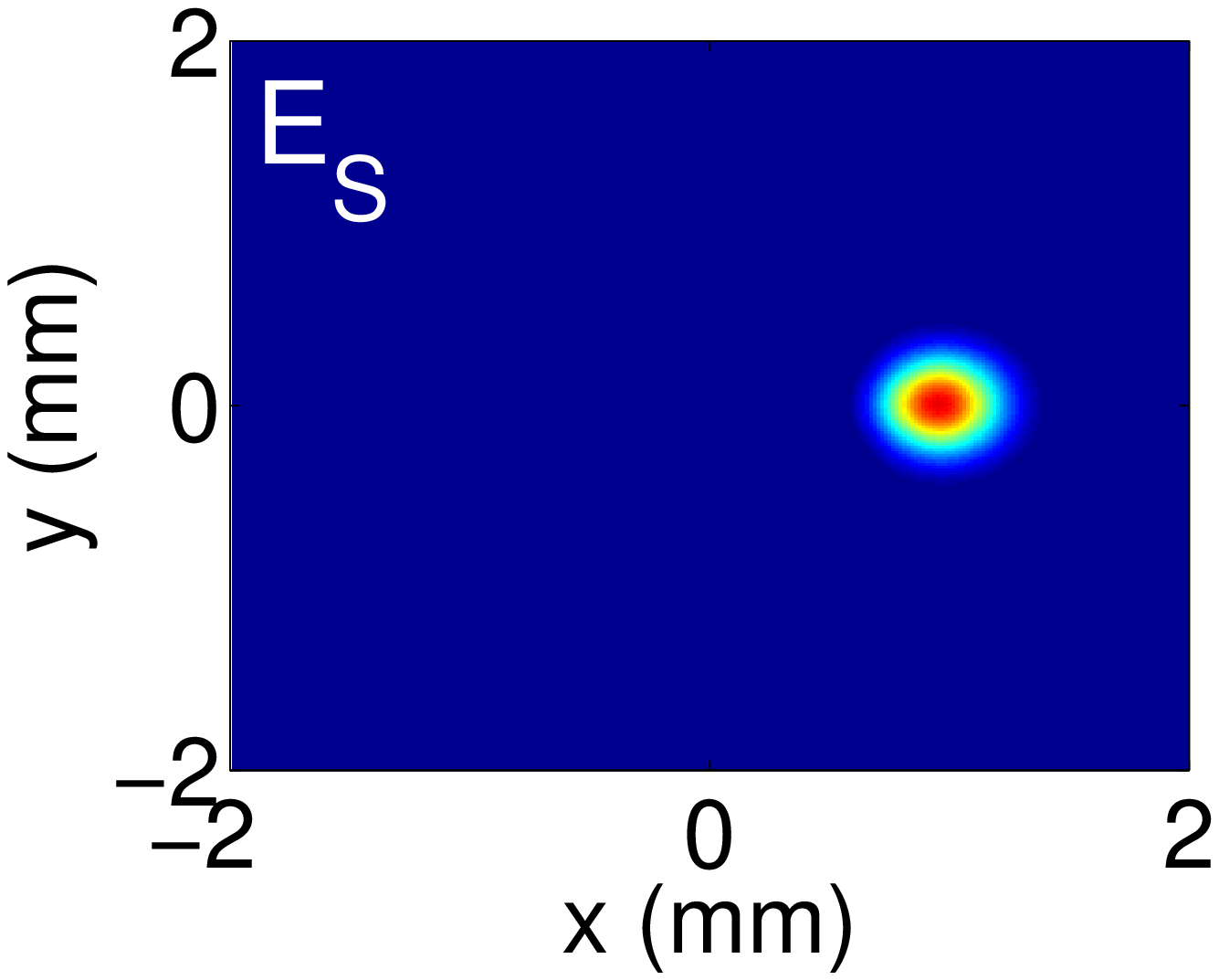}, \includegraphics[width=2.7cm]{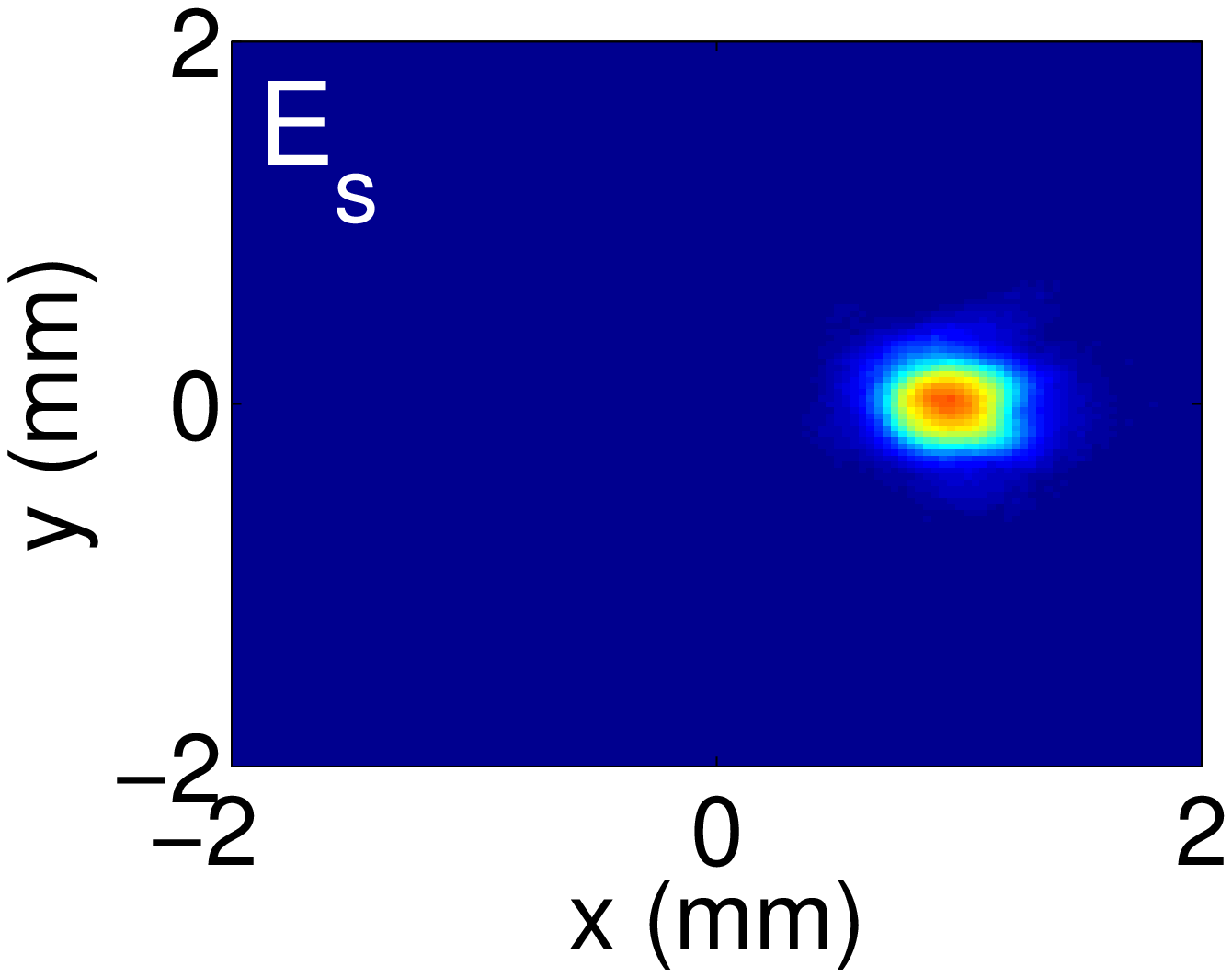} \\
         \includegraphics[width=2.7cm]{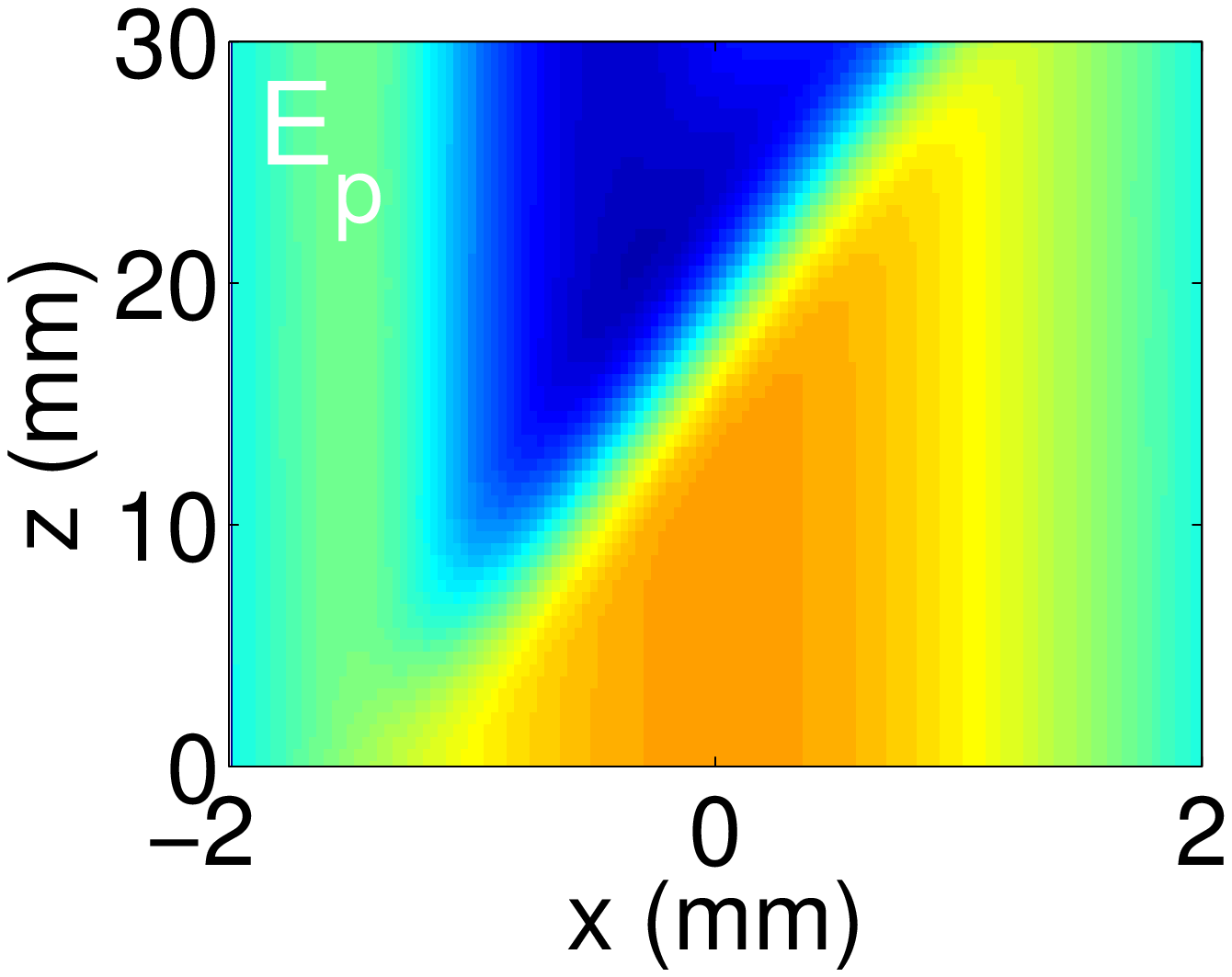}, \includegraphics[width=2.7cm]{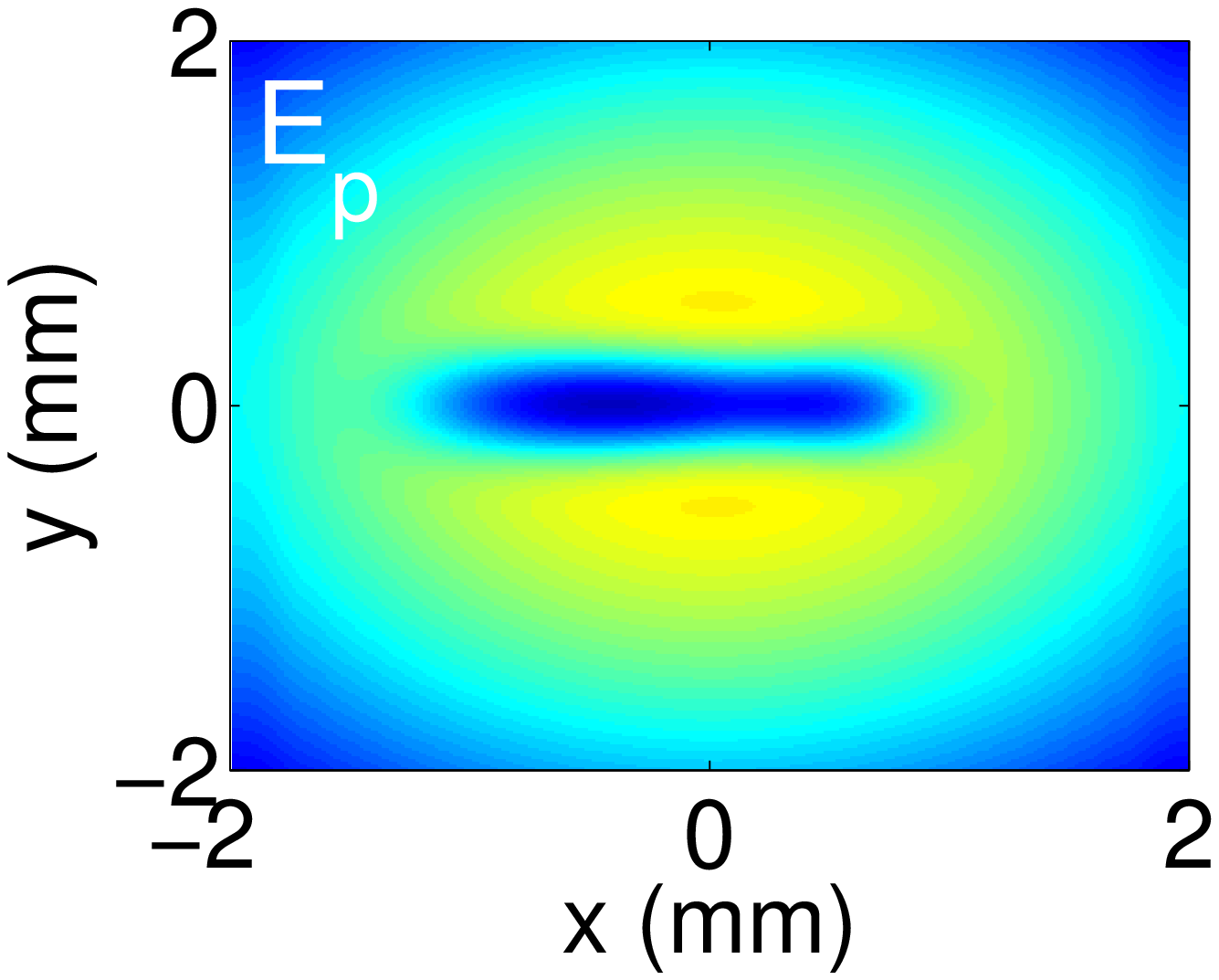}, \includegraphics[width=2.7cm]{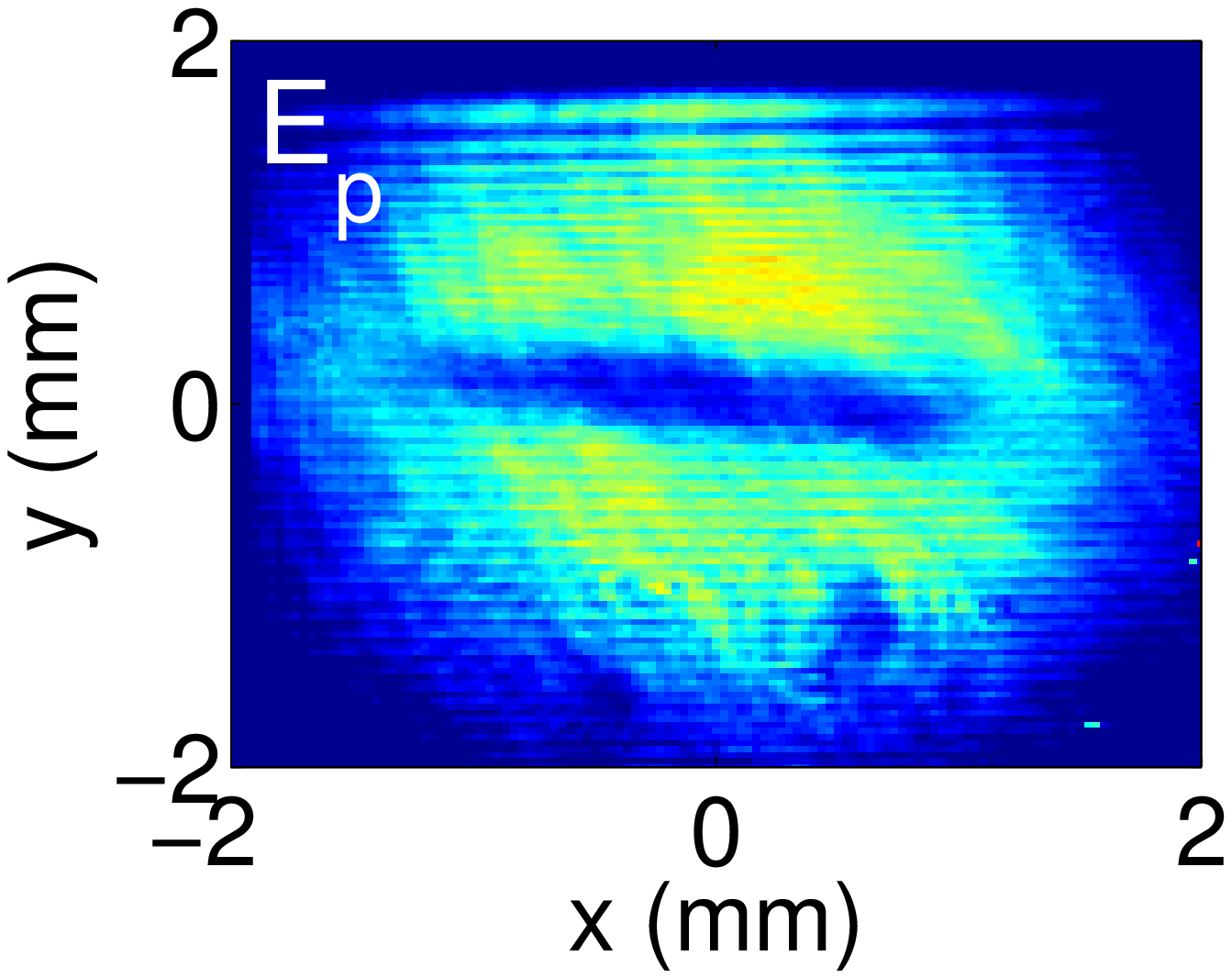} \\
     \includegraphics[width=2.7cm]{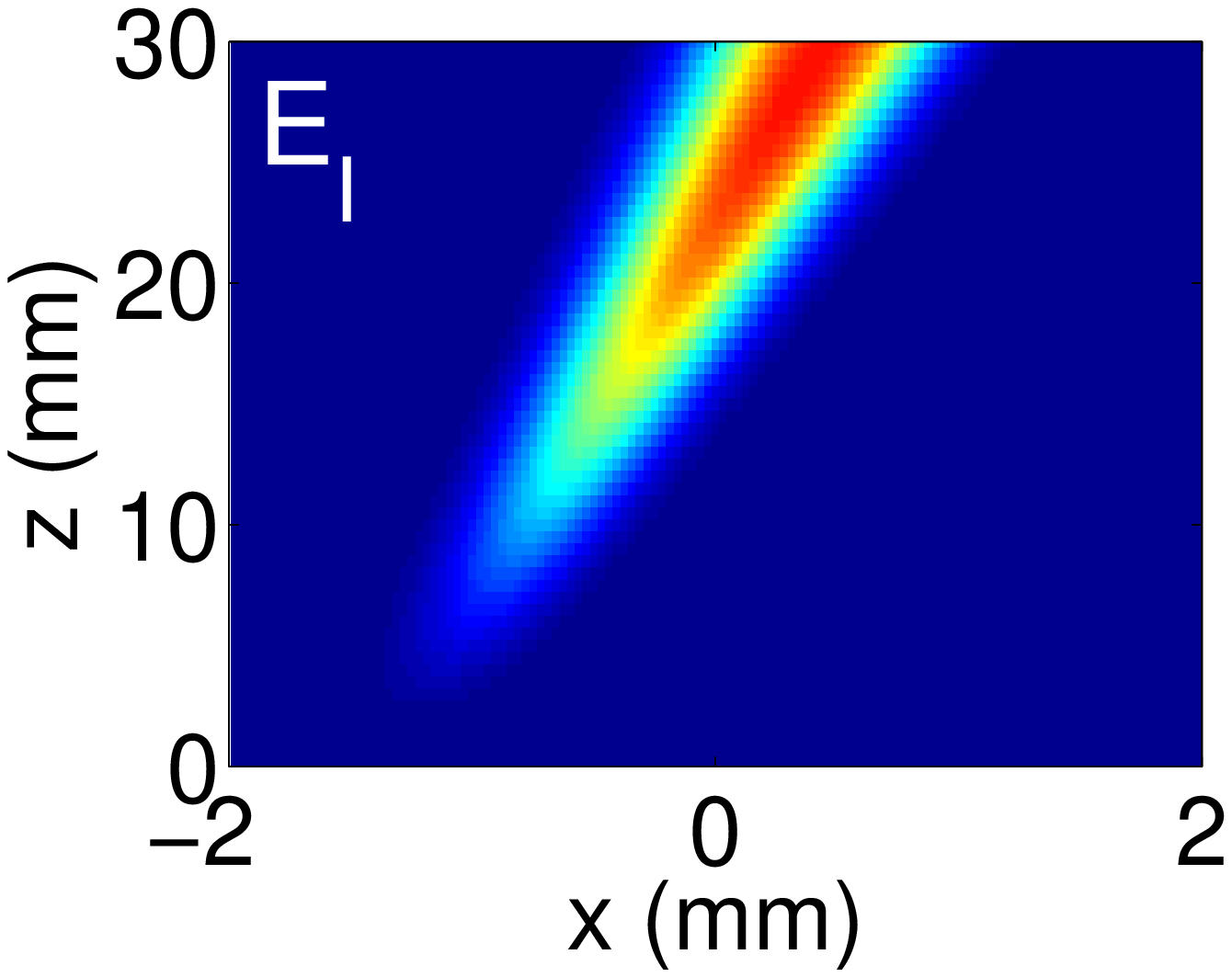}, \includegraphics[width=2.7cm]{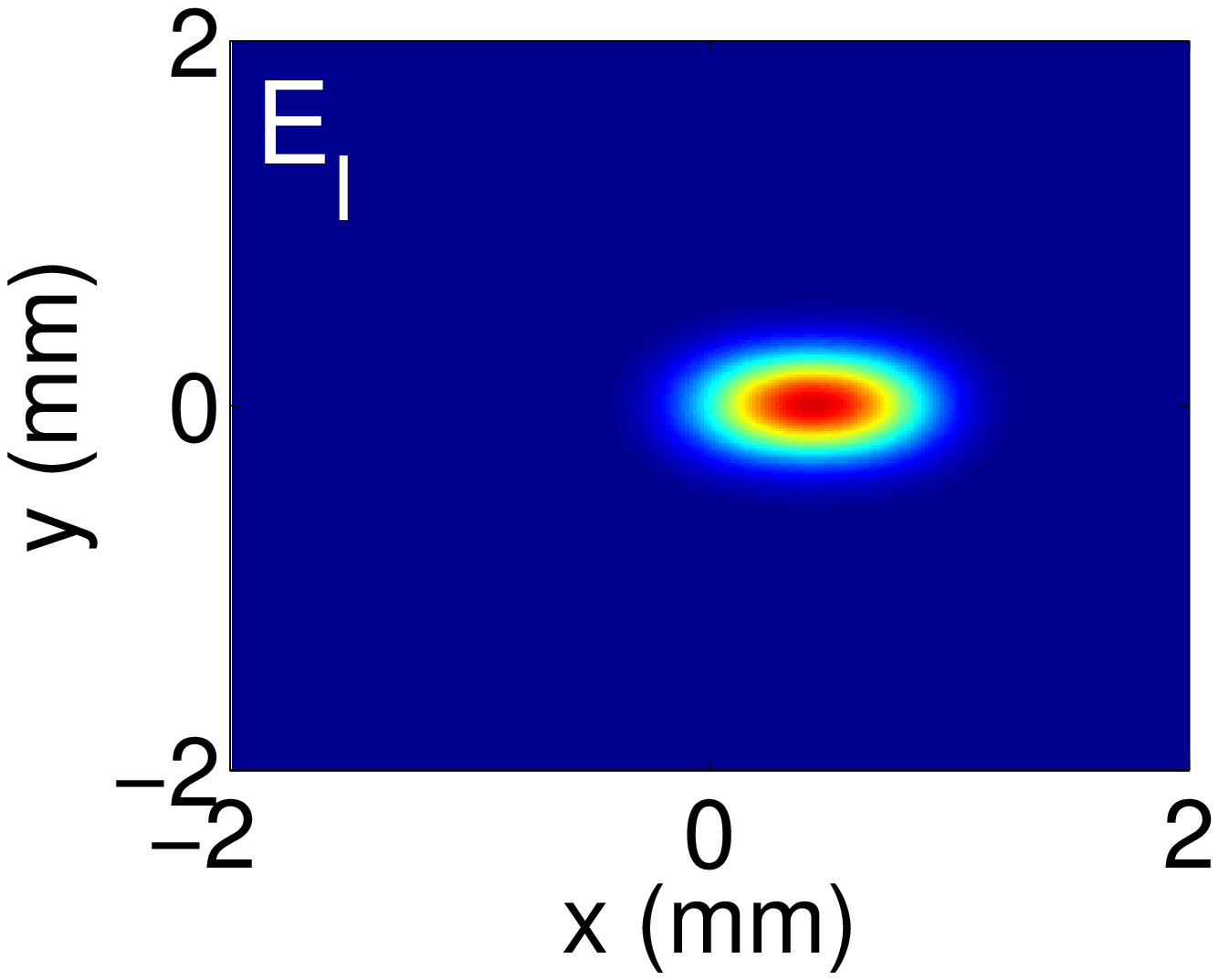}, \includegraphics[width=2.7cm]{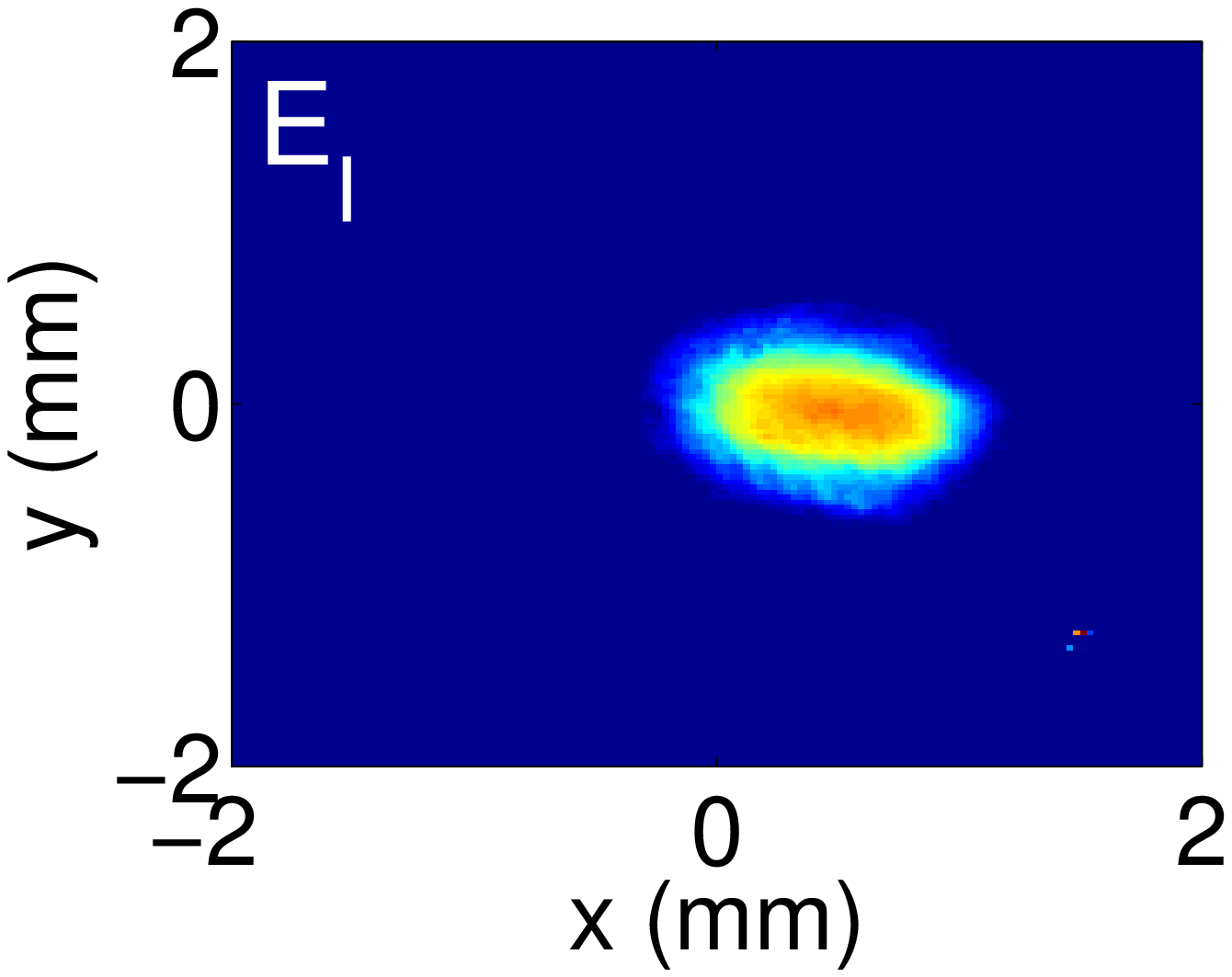}
\end{center}
     \caption{Frequency conversion. Left column, numerical dynamics of the beams $E_s$, $E_p$, $E_i$ in the
     $x-z$ ($y=0$) plane. Central column, numerical, and right column, experimental
     results at the exit face of the KTP crystal presenting
     the spatial $x-y$ output profiles of the beams.
     At the input $I_s=10 MW/cm^2$, $I_p=0.03 MW/cm^2$.
    } \label{conversione}
\end{figure}

At moderate input intensities ($I_s=10 MW/cm^2$, $I_p=0.03
MW/cm^2$), the signal interacts with the pump and an idler beam at
the second harmonic is generated (Fig. \ref{conversione}). This
regime corresponds to the well known optical non-collinear second
harmonic frequency conversion. As shown in the left column of Fig.
\ref{conversione}, the signal beam and the pump beam propagate
with their own spatial velocities (walk off angles); as long as
the signal beam overtakes the pump beam, a SH idler beam is
generated which propagates with its own characteristic spatial
linear velocity; the spatial width of the SH idler is associated
with the FF beams' interaction distance in the crystal. Signal and
pump beams are deeply depleted. Central and right columns of Fig.
\ref{conversione} report respectively the numerical and the
experimental spatial output profiles of the beams in the $x-y$
($z=L$) plane. Indeed, central and right columns of Fig.
\ref{conversione} report the existence of both the linear and the
frequency conversion regimes. In the planes parallel to the $x-z$
($y=0$) plane, low-intensity tails of the signal, along the $y$
coordinate, lead to linear beam-dynamics along the $x$ dimension
(as in the $y=1$ plane), while moderate-intensity levels of the
signal, along the $y$ coordinate, lead to frequency conversion
regime along the $x$ dimension (as in the $y=0$ plane).

\begin{figure}[h]
\begin{center}
      \includegraphics[width=2.7cm]{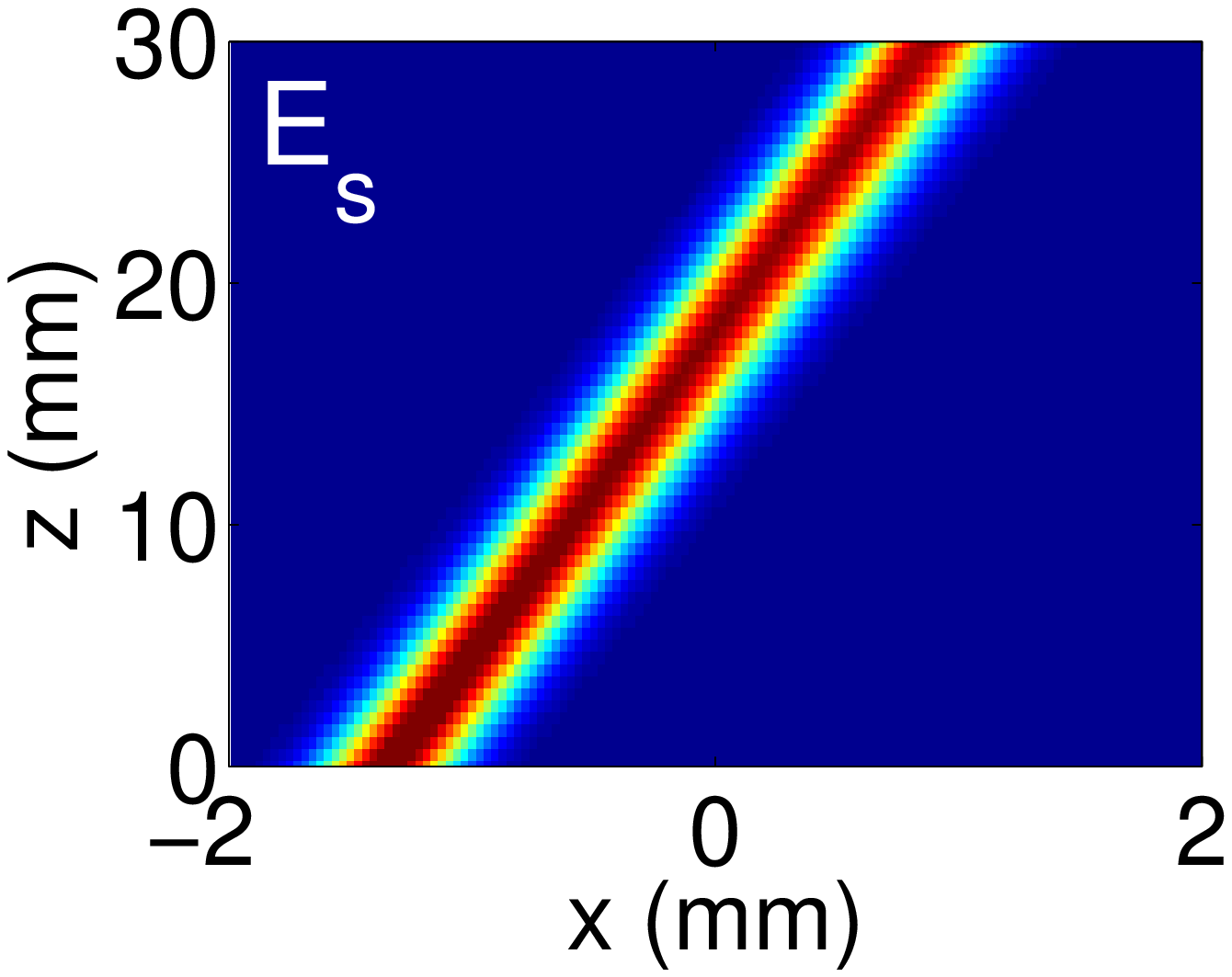},\includegraphics[width=2.7cm]{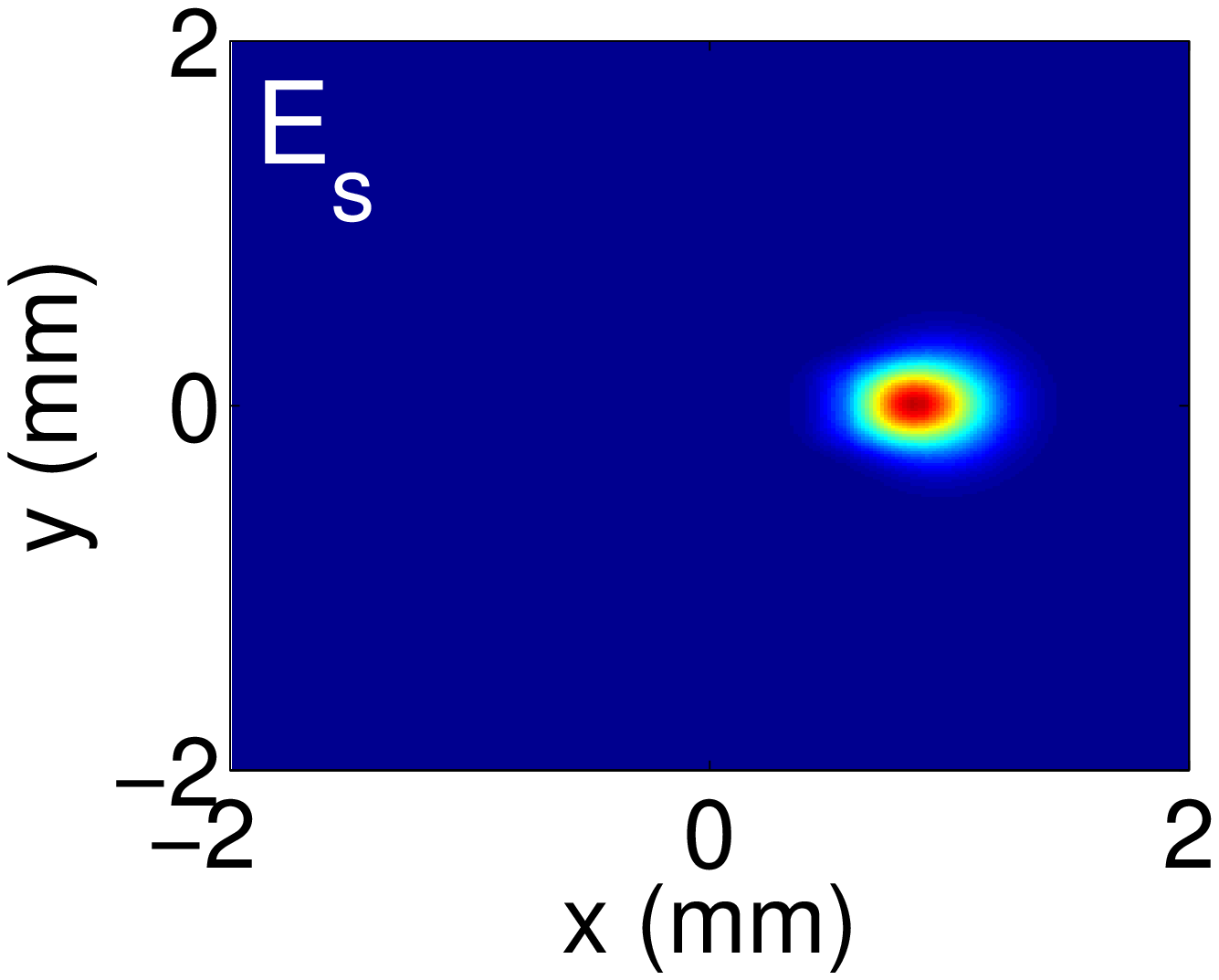}, \includegraphics[width=2.7cm]{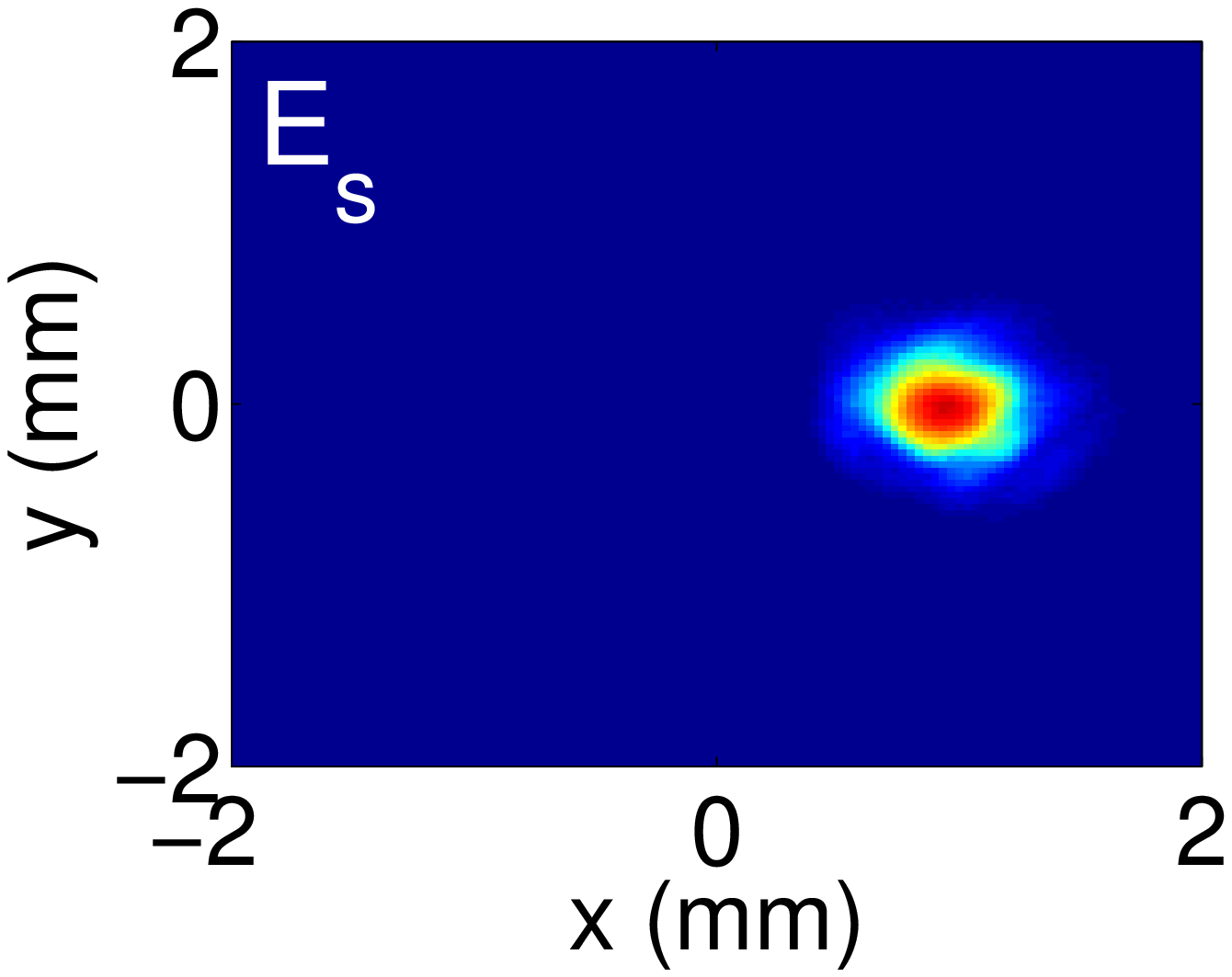} \\
       \includegraphics[width=2.7cm]{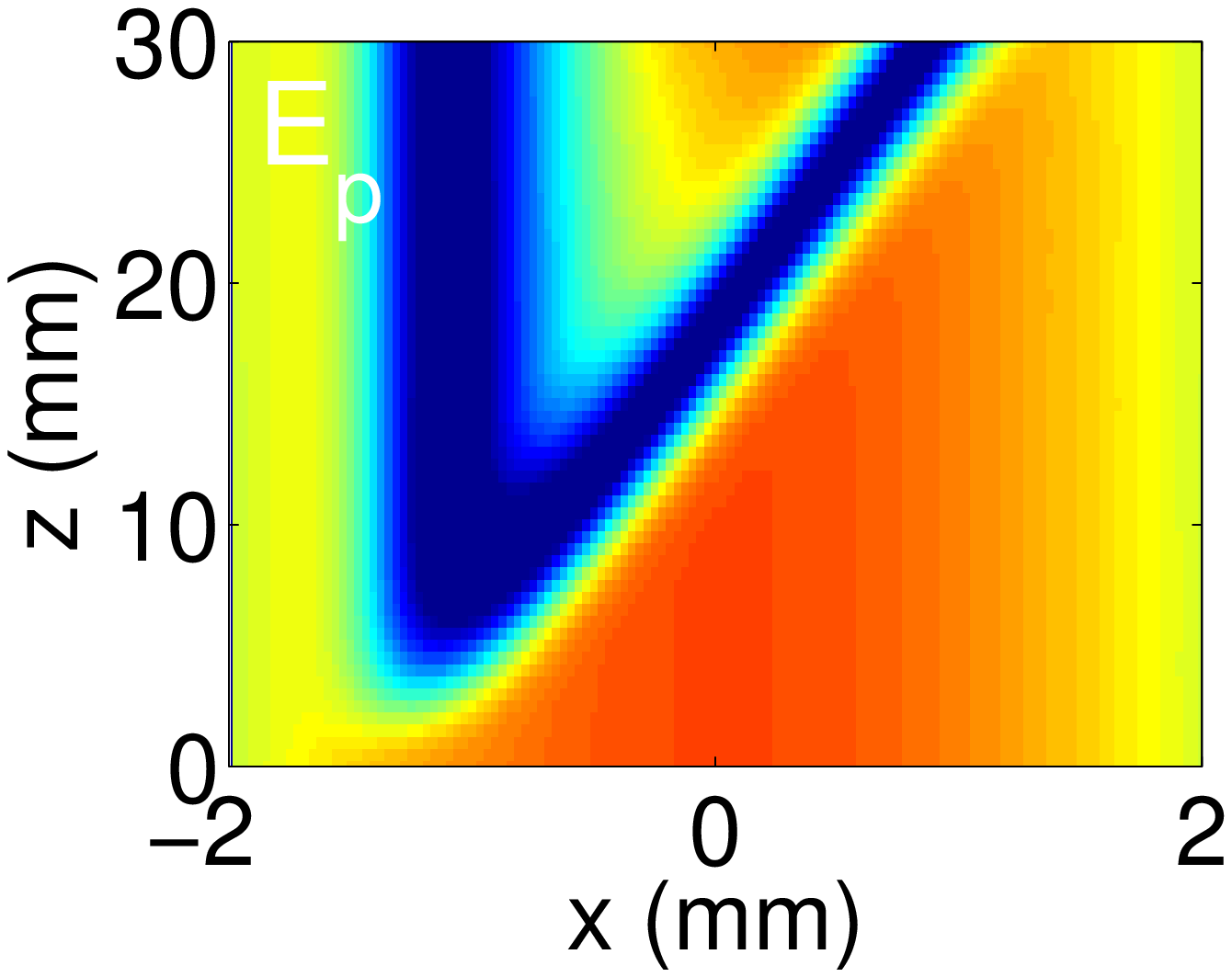},\includegraphics[width=2.7cm]{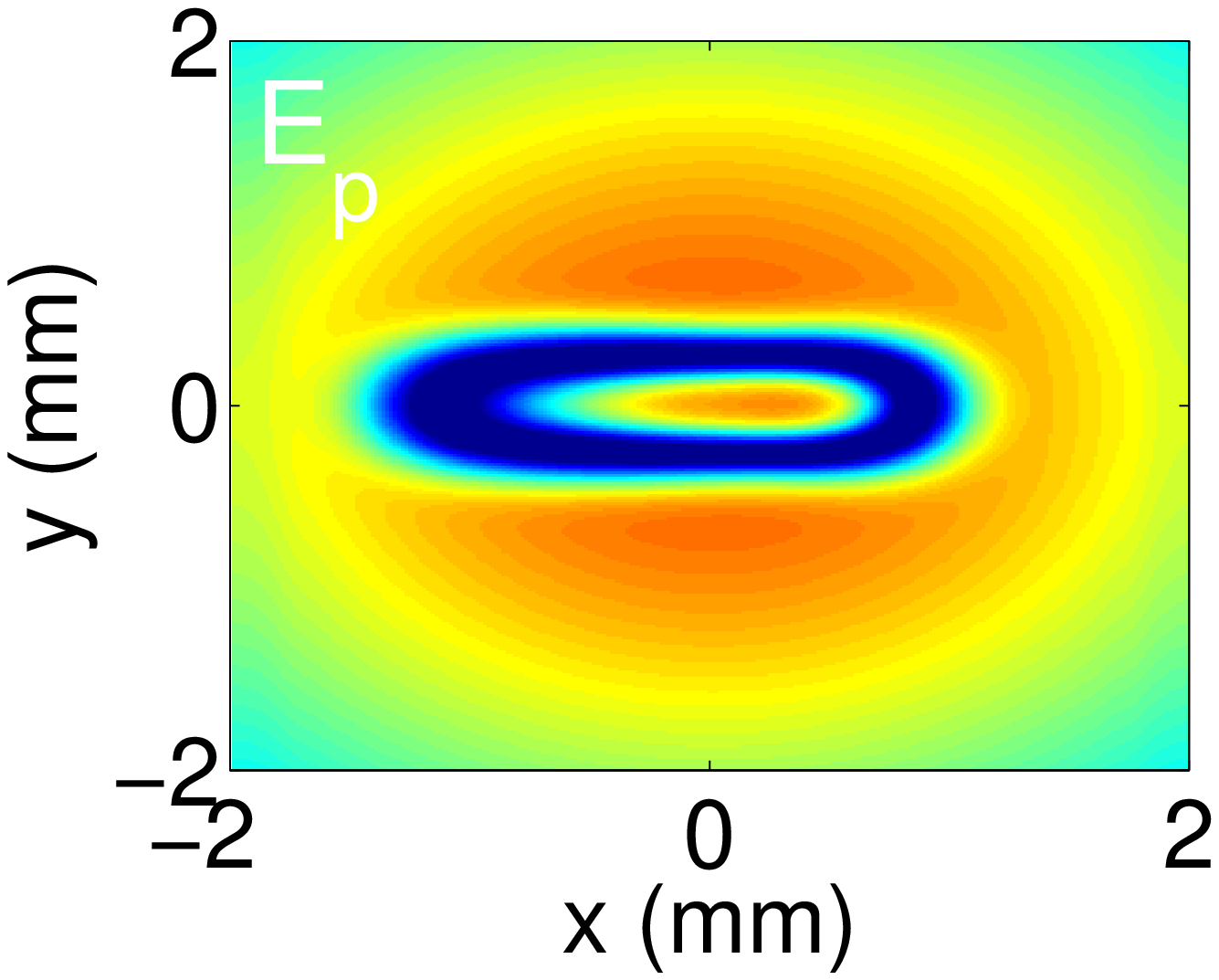}, \includegraphics[width=2.7cm]{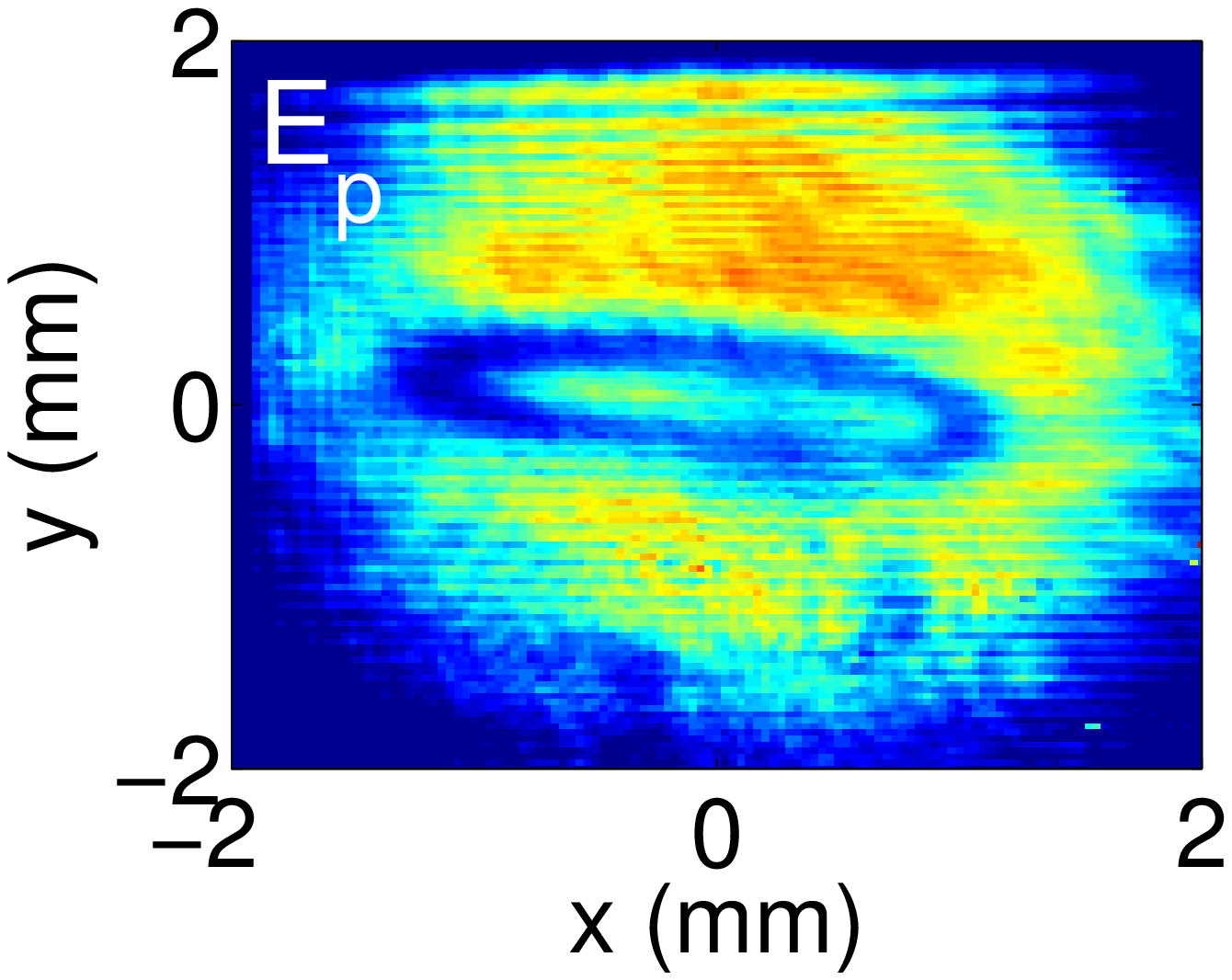} \\
       \includegraphics[width=2.7cm]{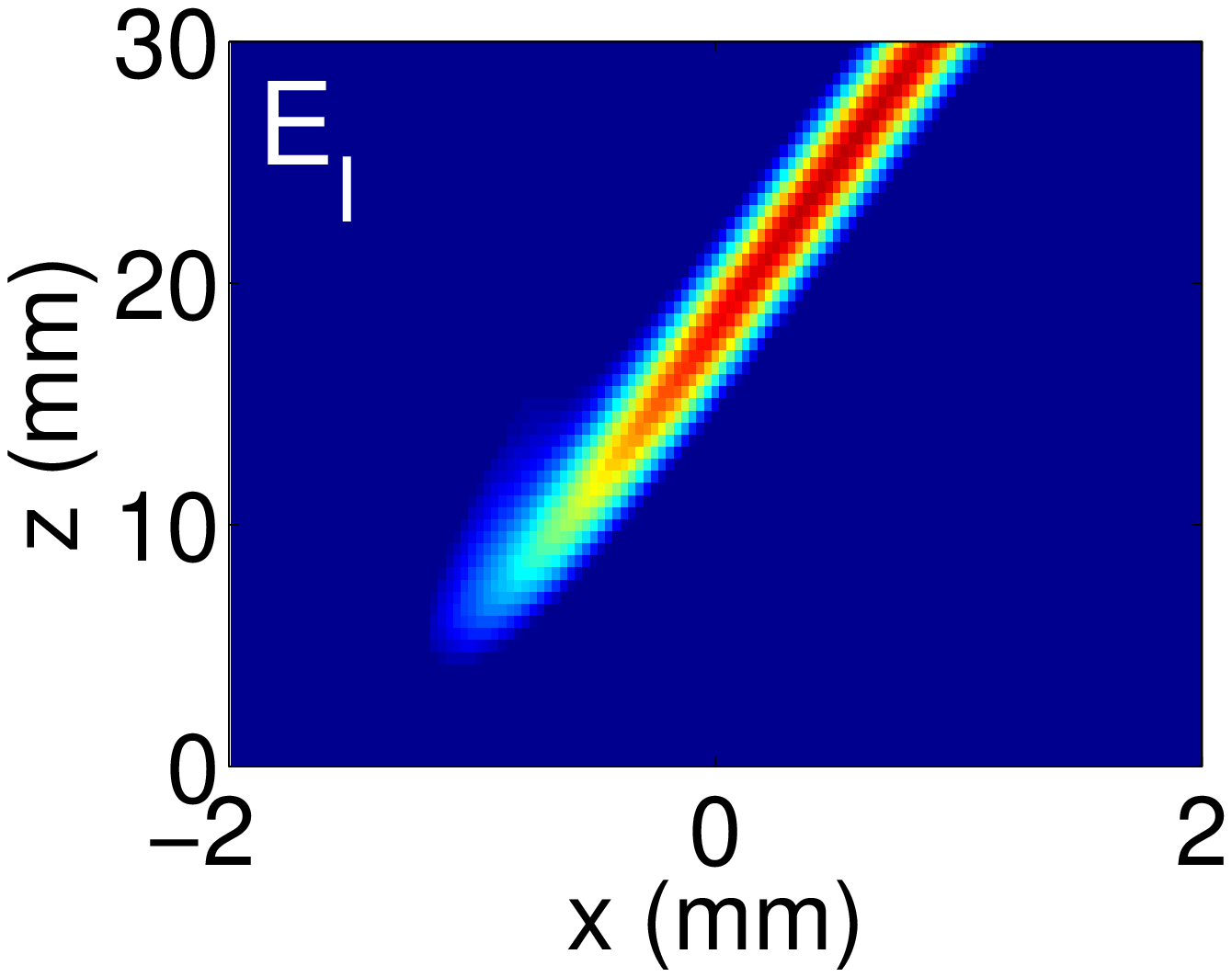},\includegraphics[width=2.7cm]{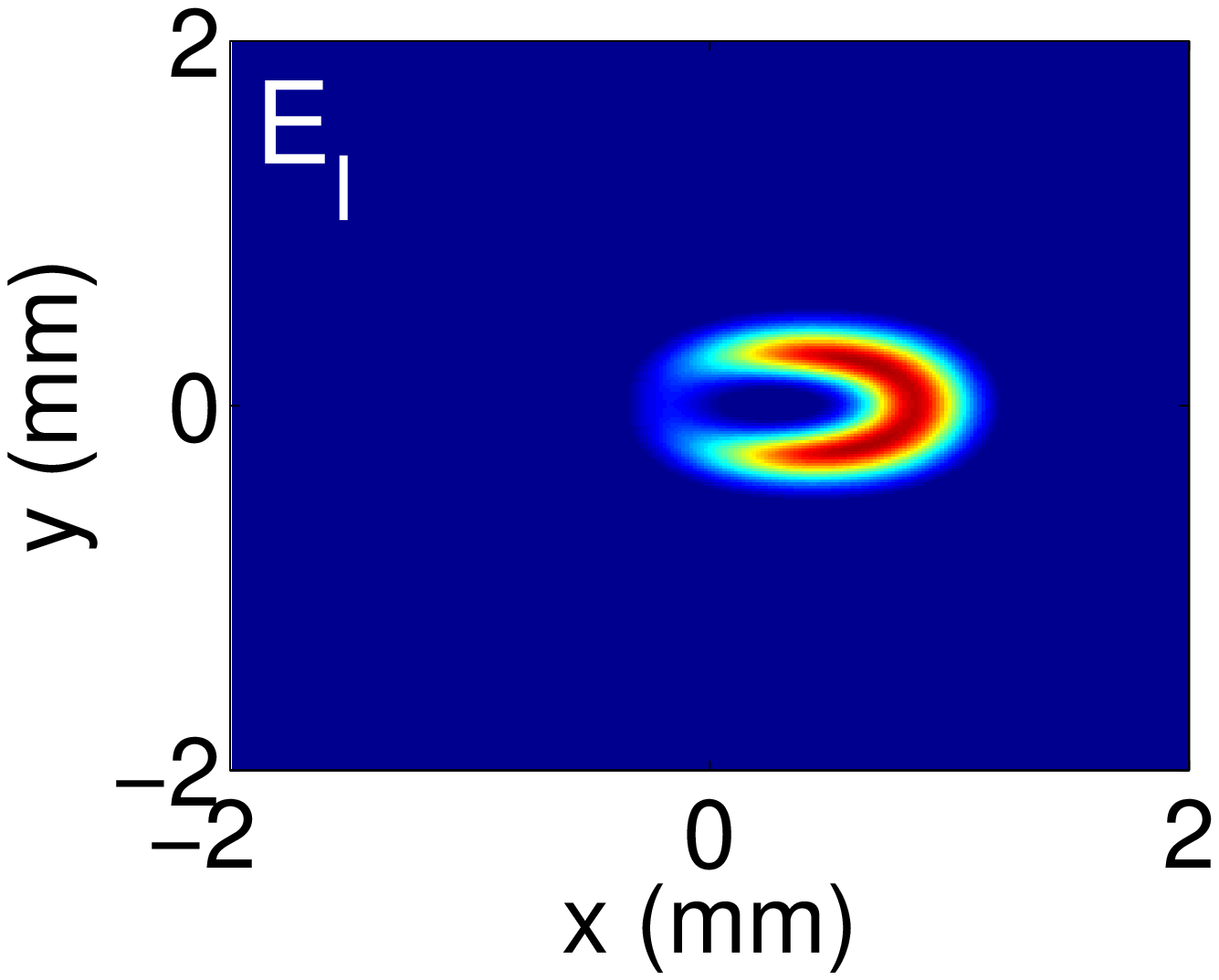}, \includegraphics[width=2.7cm]{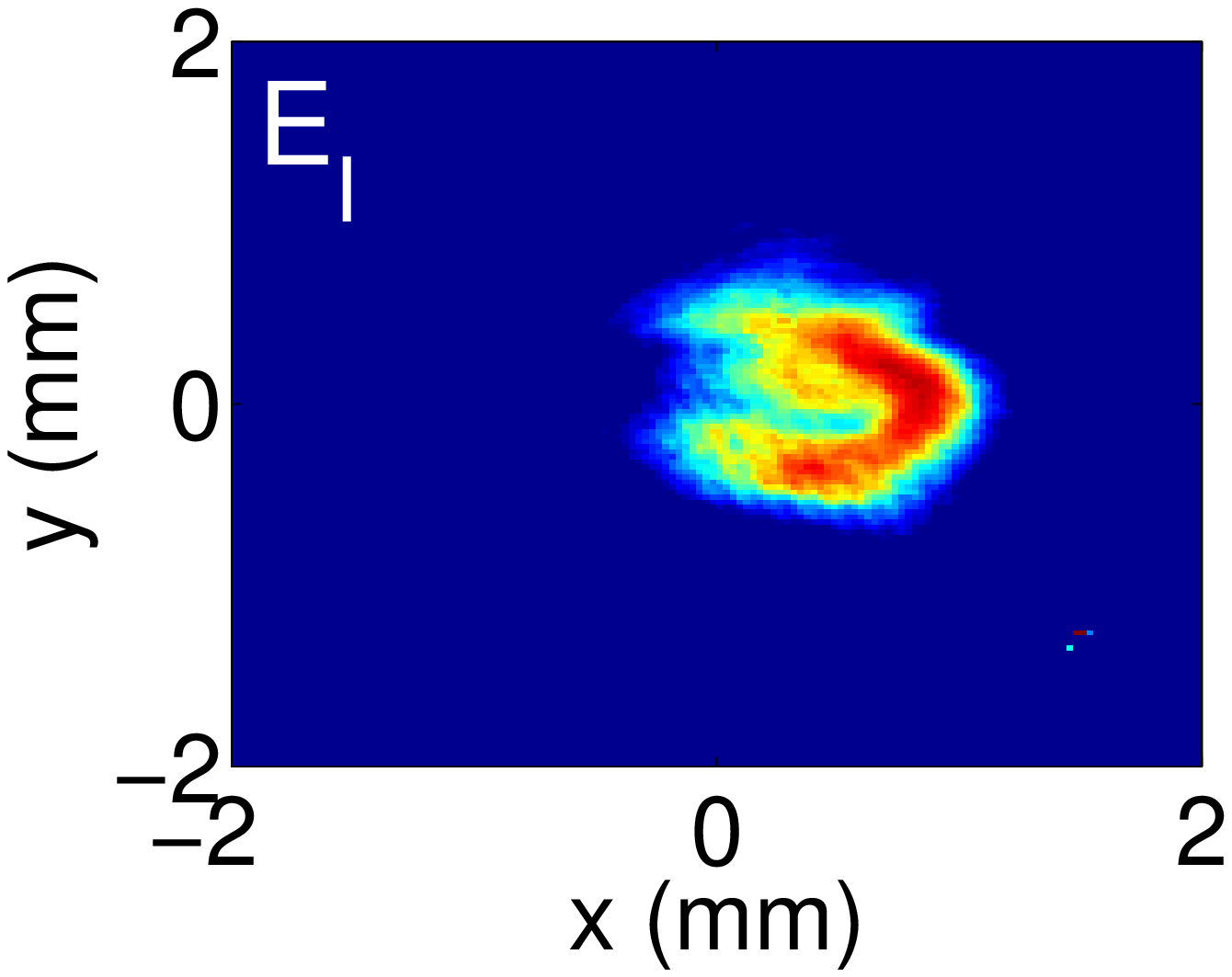}
\end{center}
     \caption{Solitary regime. Left column, numerical dynamics of the beams $E_s$, $E_p$, $E_i$ in the
     $x-z$ ($y=0$) plane. Central column, numerical, and right column, experimental
     results at the exit face of the KTP crystal presenting
     the spatial $x-y$ output profiles of the beams.
     At the input $I_s=50 MW/cm^2$, $I_p=0.1 MW/cm^2$.
    } \label{solitary}
\end{figure}

At high input intensities ($I_s=50 MW/cm^2$, $I_p=0.1 MW/cm^2$),
the scenario changes dramatically (Fig. \ref{solitary}). As shown
in the left column of Fig. \ref{solitary} the interaction of the
signal and pump beams leads to the generation of a narrow SH beam.
Additionally, a narrow dip appears in the quasi-plane wave pump;
the intensity, width and spatial direction of the signal are
slightly modified \cite{conf07}. The signal-pump interaction
generates \textbf{a stable bright-dark-bright solitary triplet}
moving with a locked spatial nonlinear velocity (nonlinear walk
off angle) that lies in between the characteristic spatial linear
velocities of the signal and the idler \cite{dega06,conf06}. The
solitary wave results from energy exchanges between
diffractionless waves of different spatial velocities.

Central and right columns of Fig. \ref{solitary} report
respectively the numerical and the experimental spatial output
profiles of the beams in the $x-y$ ($z=L$) plane. Again, central
and right columns of Fig. \ref{solitary} report the existence of
the linear regime, the frequency conversion regime and the
solitary regime. In the planes parallel to the $x-z$ ($y=0$)
plane, low-intensity tails of the signal, along the $y$
coordinate, lead to linear beam-dynamics along the $x$ dimension
(as in the $y=1$ plane); moderate-intensity levels of the signal,
along the $y$ coordinate, lead to frequency conversion regime
along the $x$ dimension (as in the $y=0.2$ plane); while
high-intensity peaks of the signal, along the $y$ coordinate, lead
to solitary regime along the $x$ dimension (as in the $y=0$
plane).
An effective way to observe the spatial shift of the directions of
the waves between moderate and high-intensity levels is to look at
the experimental and numerical spatial output profile of the
generated $E_i$ component in Fig. \ref{solitary}. In fact, we note
an horseshoe-shape of the $E_i$ component. The lateral sides of
the horseshoe represent $E_i$ waves generated by
moderate-intensity levels of $E_s$; the generated $E_i$ waves move
with the linear walk off angle $\rho_i$. The central portions of
the horse-shoe represent $E_i$ waves generated by high-intensity
levels of $E_s$; the $E_i$ waves move with nonlinear walk off
angles within the range $[\rho_i,\rho_s]$. The nearest to the
centre the highest the nonlinear walk off.

By increasing or decreasing the signal and/or pump intensities we
can observe that stable solitary triplet with different nonlinear
spatial velocity, width and energy distributions may be excited.

Note that the case reported in this paper is completely different
from the quadratic walking solitons already discussed in the
literature in the presence of non negligible diffraction
\cite{torner95,torner96}. In the same way, \textbf{our triplets
are completely different from Manakov-type vector solitons}
\cite{bart92,kang96}, where \textbf{a cubic nonlinearity} balances
diffraction.

%
%

In summary, we have shown the existence of \textbf{optical
solitary waves sustained by phase-matched nondegenerate}
three-wave parametric interaction in a quadratic KTP medium. These
solitary waves, predicted in the 1970's, are stable
velocity-locked bright-dark-bright spatial triplets, determined by
the balance between the energy exchange rates and the velocity
mismatch between the interacting waves. It is interesting to
notice that the three-wave solitary triplet concept may be applied
to describe the interaction between either three beams in the
spatial domain (diffractionless solitary waves) or three optical
pulses in the time domain (dispersionless solitary waves).

The present research in Brescia is supported by the MIUR project
PRIN 2007--CT355C.

\


\begin{thebibliography}{}

\bibitem{kara74}
Y.N. Karamzin and A.P. Sukhorukov, JEPT Lett. {\bf 20,} 339
(1974); Sov. Phys. JEPT {\bf 41,} 414 (1976).

\bibitem{tril02}
A. Buryak, P. Di Trapani, D. Skryabin, and S. Trillo, Phys. Rep.
{\bf 370,} 63 (2002).


\bibitem{armstrong70}
J. A. Armstrong, S. S. Jha, and N. S. Shiren, IEEE J. Quantum
Electron. {\bf QE-6,} 123 (1970).

\bibitem{nozaki73}
K. Nozaki and T. Taniuti, J. Phys. Soc. Jpn. {\bf 34,} 796 (1973).

\bibitem{kaup79}
D. J. Kaup, A. Reiman, and A. Bers, Rev. Mod. Phys. {\bf 51,} 275
(1979).

\bibitem{tril96} S. Trillo, Opt. Lett. {\bf 21,} 1111 (1996).

\bibitem{dega06} A. Degasperis, M. Conforti, F. Baronio, and S. Wabnitz,
Phys. Rev. Lett. {\bf 97,} 093901 (2006).


\bibitem{mcca67} S. McCall and E. Hahn, Phys. Rev. Lett. {\bf 18,} 908 (1967).

\bibitem{druh83} K. Druhl, R. Wenzel, and J. Carlsten, Phys.
Rev. Lett. {\bf 51,} 1171 (1983).

\bibitem{russ09} A. Abdolvand, A. Nazarkin, A. Chugreev, C. Kaminski, and P. Russel, Phys.
Rev. Lett. {\bf 103,} 183902 (2009).

\bibitem{pich91} E. Picholle, C Montes, C. Leycuras, O. Legrand,
and J. Botineau, Phys. Rev. Lett. {\bf 66,} 1454 (1991).


\bibitem{conf07}
M. Conforti, F. Baronio, A. Degasperis, and S. Wabnitz, Opt.
Express {\bf 15,} 12246 (2007).

\bibitem{conf06}
M. Conforti, F. Baronio, A. Degasperis, and S. Wabnitz, Phys. Rev.
E {\bf 74,} 065602(R) (2006).

\bibitem{torner95}
L. Torner, W. Torruellas, G. Stegeman, and C. Menyuk, Opt. Lett.
{\bf 20,} 1952 (1995).

\bibitem{torner96}
W. Torruellas, G. Assanto, B. Lawrence, R. Fuerst, and G.
Stegeman, Appl. Phys. Lett. {\bf 68,} 1449 (1996).


\bibitem{bart92}
M. Shalaby and A. Barthelemy, IEEE J. Quantum Electron. {\bf 28,}
2736 (1992).

\bibitem{kang96}
J. U. Kang, G. I. Stegeman, J. S. Aitchison, and N. Akhmediev,
Phys. Rev. Lett. {\bf 76,} 3699 (1996).




%
%
%
%
%
%
%
%
%
%
%
%
%
%
%





%
%
%
%
%
%
%
%
%
%
%
%
%
%
%
%
%
%
%
%
%
%
%
%
%
%
%
%
%
%
%
%
%
%
%
%
%
%
%






\end{thebibliography}
\end{document}